\documentclass[num-refs]{wiley-article}
\usepackage{siunitx}
\usepackage{hyperref}
\usepackage{algorithm}
\usepackage{algpseudocode}
\usepackage{longtable}
\usepackage{float}
\usepackage{ulem}
\usepackage{footnote} 
\usepackage{xcolor}

\usepackage{siunitx}

\papertype{Original Article}
\paperfield{Journal Section}

\title{Addressing Positivity Violations in Extending Inference to a Target Population}

\author[1\authfn{1}]{Jun Lu MS}
\author[1\authfn{2}]{Sanjib Basu PhD}

\affil[1]{Biostatistics, School of Public Health, University of Illinois Chicago, Chicago, IL, USA}

\corraddress{School of Public Health, 1603 W. Taylor St. | 955 SPHPI| Chicago, IL 60612, USA}
\corremail{sbasu@uic.edu}

\runningauthor{Lu and Basu}

\begin{document}

\begin{frontmatter}
\maketitle

\begin{abstract}
Enhancing the external validity of study results is essential for their applicability to real-world populations. However, if the positivity assumption is violated because some subgroups are underrepresented or absent from the study, the external validity of these findings may be compromised. To address positivity violations in estimating the average treatment effect for a target population, we propose a framework that integrates characterizing the un/underrepresented group and performing sensitivity analysis for inference in the original target population. Our approach helps identify limitations in study sampling and improves the robustness of study findings for real-world populations. We apply this approach to extend findings from phase IV trials of treatments for opioid use disorder to a real-world population based on the 2021 Treatment Episode Data Set.

\end{abstract}

\end{frontmatter}

\newpage

\section{Introduction}
Ensuring that the study findings are relevant and applicable in real-world settings is essential to improve treatment effectiveness and inform practical decision making. However, the gap between study samples and the broader real-world population can compromise the wider applicability of the findings. In practice, this gap often arises from various factors influencing study participation, such as restrictive inclusion criteria, limited geographic diversity, and socio-demographic biases \cite{kim2017broadening}. For instance, Black women—who have a 40\% higher mortality rate—have been underrepresented in breast cancer clinical trials over the past two decades, raising concerns about the applicability of trial results to diverse populations \cite{nguyen2023race}. The gap between study results and their broader applicability has garnered increasing attention, particularly in light of the United States Food and Drug Administration's guidance document \textit{Enhancing the Diversity of Clinical Trial Populations — Eligibility Criteria, Enrollment Practices, and Trial Designs} \cite{FDA2019EnhancingDiversity}. This guidance emphasizes the need to broaden eligibility criteria to improve the generalizability of trial results and ensure they better reflect the populations likely to use the therapy if approved.

To enhance external validity for the target population, two designs—nested and non-nested—and a range of corresponding methods have been developed. In the nested design, investigators sample a large cohort sample representative of the real-world population and then select a study sample, a subset of the cohort, to follow for the outcome of interest, with the goal of generalizing findings from the study sample to the cohort sample. In contrast, the non-nested design employs independent sampling, where a study sample is selected for outcome observation while an external sample is obtained separately to represent the target population, allowing for the transportation of findings from the study sample to the external sample. These methods for nested and non-nested designs are reviewed in \citet{ling2023overview}. In particular, \citet{dahabreh2019generalizing} reviewed outcome regression and inverse propensity weighting methods and proposed doubly robust estimators to generalize nested trial findings to all trial-eligible patients within a large cohort. Later, \citet{dahabreh2020extending} extended this work to transport the trial findings to the trial nonparticipants in nested and nonnested trial designs.

Methods to improve external validity typically rely on two key assumptions: conditional exchangeability and the positivity of study participation. Conditional exchangeability requires that, when conditioned on observed covariates, the potential outcome is independent of study participation. The positivity of study participation requires that, given the observed covariates in the target population, every individual has a probability of being included in the study that is bounded away from 0.  These two key assumptions are frequently violated in real-world settings, making applications of such methods challenging. To address violations of conditional exchangeability, especially when unobserved covariates confound the relationship between study participation and potential outcomes, a range of sensitivity analysis techniques have been developed \cite{lesko2016effect, nguyen2017sensitivity, nguyen2018sensitivity, steingrimsson2023sensitivity, dahabreh2023sensitivity, huang2024sensitivity}. Furthermore, \citet{nilsson2023proxy} proposed using proxy variables to handle this issue. In contrast, literature addressing violations of study participation assumptions is limited. Recently, \citet{chen2023generalizability} introduced a generalizability score that combines participation and propensity scores to identify well-represented subpopulations, while \citet{parikh2024we} proposed an optimization-based framework for this purpose. Similarly, \citet{huang2024overlap} developed a sensitivity framework to address external validity under overlap violations by decomposing bias into three parameters, and \citet{zivich2023synthesis, zivich2024transportability} advocated using mathematical models and synthesis estimators to address positivity violations caused by a single continuous or binary variable. Despite these advancements, a unified framework that integrates positivity checks with inference under assumption violations remains underdeveloped.

When positivity assumptions are violated, researchers typically focus on two challenging questions: (1) Which subjects, and what proportion of the target population, can not be reliably estimated from the current study sample? (2) What potential bias arises when these subjects are excluded or their outcomes are inferred? To address these questions, we propose a novel framework that tackles positivity violations when making inferences about a target population. In brief, the target population is divided into three groups based on the study sample and the external target sample: (1) an unrepresented group, (2) an underrepresented group, and (3) a well-represented group. Common weighting estimators are applied to estimate the average treatment effects (ATE) for the well-represented group, while simple sensitivity analysis parameters are introduced to estimate the ATE for the original target population. This framework helps researchers identify which parts of the population are underrepresented in the study, enabling them to report the limitations of study recruitment. It also provides a way to assess the robustness of the study conclusions by incorporating sensitivity analysis.

The remainder of this article is organized as follows: Section 2 reviews methods for external validity, including notation, assumptions, weighting estimators, and establishes their efficiency bounds in the non-nested design. The efficiency bound illustrates how study sampling impacts the efficiency of inference in the target population. Section 3 presents the proposed method for addressing positivity violations. Section 4 evaluates the performance of the proposed method through simulations. Section 5 applies the framework to a real-world example, extending inferences from a phase IV trial for treating opioid use disorder to a real-world population based on the 2021 Treatment Episode Data Set. Finally, Section 6 concludes with a discussion. An R package (https://github.com/junlu1995/TransPos) was developed to implement the proposed framework.

\section{Methods for Transportability}
\subsection {Notation and Assumption}
In this article, we focus on the non-nested design, in which the study sample and the external target sample are obtained independently \citep{dahabreh2021study}. Our method can also be easily extended to nested designs, where a subset of participants from a larger observational cohort is selected for study \cite{olschewski1985comprehensive}. The objective is to evaluate the comparative effectiveness of treatments for the target population.

We consider the framework of a superpopulation with biased sampling to account for the non-nested design mechanism \cite{jewell1985least}. Specifically, a sample of size \(\tilde{n}\) is first drawn from an infinite population, and separated into a sample of size \(n_1\) study sample and a sample of size \(\tilde{n} - n_1\) non-study sample. Then a subset of a sample of size \(n_2\) non study sample is selected as the external target sample. Sample size and covariate information are only available for the study sample and the external target sample, while additional outcome information is obtained for the study sample.

\begin{figure}[!h]
    \centering
    \includegraphics[width=0.90\textwidth]{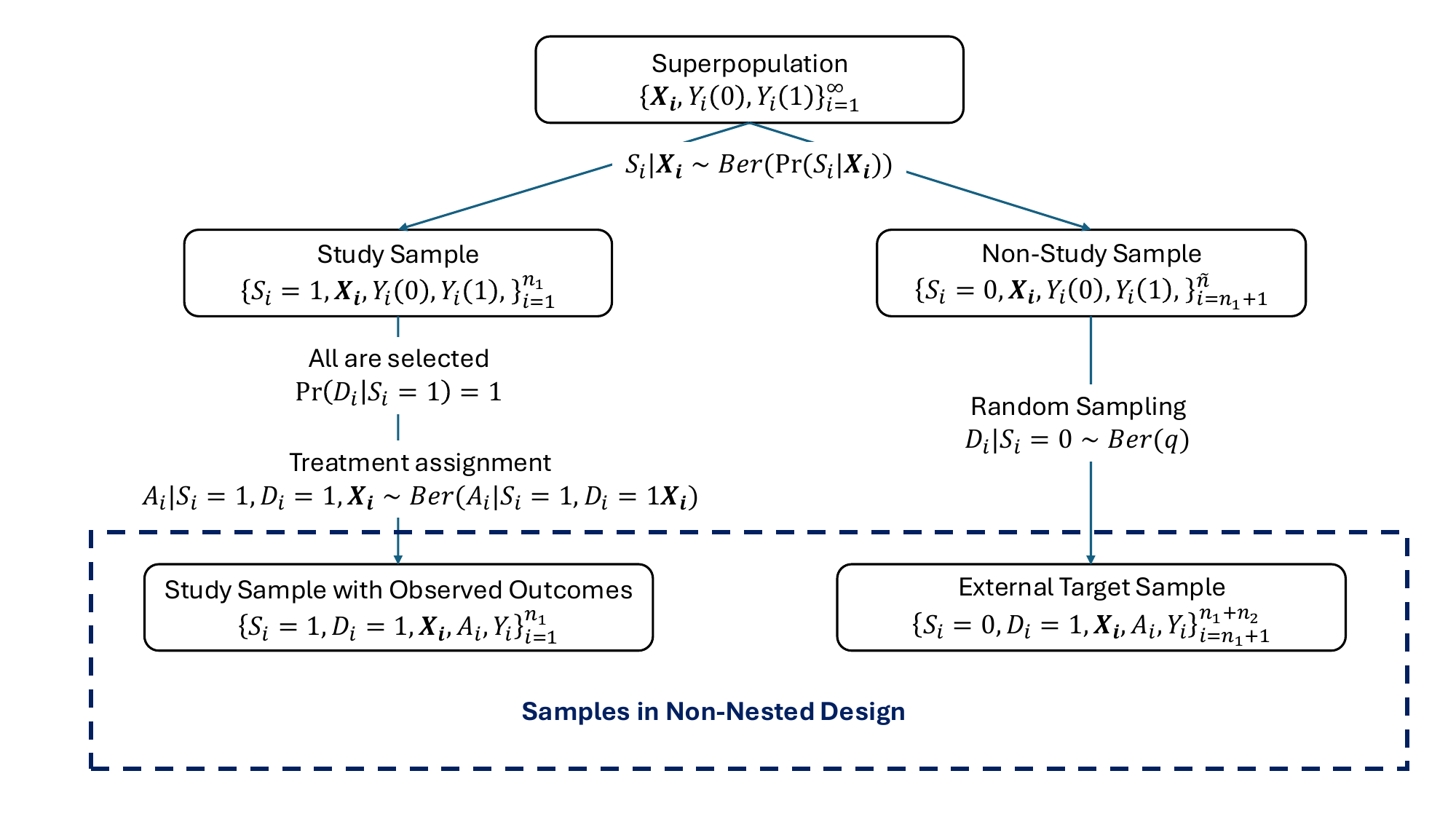}
    \caption{Demonstration of the study sampling, treatment assignment, and biased sampling in the non-nested design under a superpopulation framework.}
    \label{fig:fig1}
\end{figure}

For each unit $i$, let $S_i$ be a binary indicator of study participation ($S_i = 1$ for study sample, $S_i = 0$ for non study sample); $D_i$ be a binary indicator for selection into the non-nested study ($D_i = 1$ for selected, $D_i = 0$ for not selected); $A_i$ a binary treatment indicator ($A_i = 1$ for treated, $A_i = 0$ for untreated); $\bm{X_i}$ a $q$-dimensional vector of pre-treatment covariates; $Y_i$ the observed outcome; and $Y_i(a)$ the potential outcome under treatment $a$. The observable data $O_i$ consist of $\{S_i = 1, D_i = 1, \bm{X_i}, Y_i, A_i\}$ for study sample ($i = 1, \ldots, n_1$) and $\{S_i = 0, D_i = 1, \bm{X_i}\}$ for external target sample ($i = n_1 + 1, \ldots, n$). The data structure is shown in Figure \ref{fig:fig1}.

Individuals participate in the study with probability \( e_s = \Pr(S=1|\bm{X}) \), and study participants are assigned to treatment \( a \) with probability \( e_a = \Pr(A=a|\bm{X}) \), where \( a = 0 \) or \( 1 \). All study participants are selected into the non-nested design with \( \Pr(D=1|S=1) = 1 \), while only a random subset of study nonparticipants is selected, with unknown equal probability \( \Pr(D=1|S=0) = q \). In addition, in the biased sampling model, since \( e_s \) is not identifiable \cite{dahabreh2020extending}, we are interested in the conditional probability of study participation given selection for the non-nested design, defined as \( h_s = \Pr(S=1|\bm{X}, D=1) \).

Our estimand of interest is the ATE of the target nonparticipants, given by
$$\tau = \mathbb{E}(Y(1)- Y(0)|S= 0)$$

 The following assumptions are commonly made for identification of $\tau$ (omitting the subscripts):
\begin{itemize}
    \item[Conditional Treatment Exchangeability:]
    It assumes that, conditional on covariates \(\bm{X}\), the treatment assignment \(A\) is independent of the potential outcomes \(Y(1)\) and \(Y(0)\). That is $Y(1), Y(0) \perp A \mid \bm X, S = 1$.
    \item[Positivity of Treatment Assignment]:
    It requires that for all values of \(\bm X\) in the target population, there is a probability of receiving either treatment or control bounding away from 0. That is, there exists a constant $c$ such that $Pr(A = a|\bm X = \bm{x}, S = 1) > c$ for every $a$ and every $x$ with positive density $f_{\bm X|S = 0}(\bm x) > 0$.
    \item[Conditional Exchangeability for Study Participation] :
    It assumes that, conditional on covariates \(\bm X\), the study participation is independent of the potential outcomes \(Y(1)\) and \(Y(0)\). That is  $Y(1), Y(0) \perp S \mid \bm X$.
    \item[Positivity of Trial Participation]:
    It requires that for all values of the $\bm X$ in the target population, the probability of participating in the trial and observing the outcome is bounded away from 0. That is, there exists a constant $c$ such that $Pr(S = 1|\bm X = \bm{x}) > c$ for every $x$ with positive density $f_{\bm X|S = 0}(\bm x) > 0$.
    \item[Stable Unit Treatment Value Assumption (SUTVA) for Trial Participation]: It assumes no interference between subjects who are selected for the study and those who are not. It also requires treatment version irrelevance between the study and target samples, ensuring consistent treatment across both populations. If $S = 1, A = a$, then $Y = Y(a)$.
    \item[Random Biased Sampling]: It assumes all study participants are included in the non-nested design, while only a random subset of the study non-participants are included. That is $
   Pr(D=1|S=1) = 1
    $, $Pr(D=1|S=0) = Pr(D|S=0, \bm X, Y(1), Y(0))$, $D \perp (\bm X, A, Y(1), Y(0))|S$.
\end{itemize}

 \subsection{Weighting Estimator and Efficiency Bound}
Under the aforementioned assumptions, ATE $\tau$ can be identified from the observed data, and several estimators have been proposed for its estimation, as reviewed by \citet{li2022generalizing}. We consider the weighting estimator here, which adjusts for unequal probabilities of study participation and treatment assignment. This approach ensures an an unbiased estimation by giving more weight to underrepresented observations and reducing the weight to overrepresented ones. A commonly used weighting estimator is the Hájek estimator in survey sampling \cite{hajek1971comment}, given by
$$
\hat \tau_w = \frac{\sum_{i=1}^n \hat w_{1i}Y_i}{\sum_{i=1}^n \hat w_{1i}} - \frac{\sum_{i=1}^n \hat w_{0i}Y_i}{\sum_{i=1}^n \hat w_{0i}}
$$
with estimated weights $w_{ai}$ for $a = 0, 1$ calculated by 
$$ w_{a}(S, A, \bm X; \bm\alpha, \bm\beta) = I(S= 1, A = a)\frac{(1-h_s(\bm{X}; \beta))}{h_s(\bm{X}; \beta)e_a(\bm{X}; \alpha)}$$
\vspace{-10pt}
$$\hat w_{ai} = w_{a}(S_i, A_i, \bm X; \hat{\bm\alpha}, \hat{\bm\beta})$$
where $I(\cdot)$ is the indicator function; $h_s(\bm{X_i}; \bm\beta)$ is the study sampling score conditional on non-nested design selection modeled by a generalized linear model, and $e_a(\bm{X_i}; \bm\alpha)$ is the propensity score modeled by a generalized linear model. Although $Y$ for $n_1+1,...,n$ are not observed, their weights are 0s removing their effect in the summation. After applying weighting, the covariates are expected to be balanced between the treatment and control groups and comparable to the distribution of the target population. When both $h_s(\bm{X_i}, \bm\beta)$ and $e_a(\bm{X_i}, \bm\alpha)$ are correctly specified, $\hat\tau_w$ is consistent for $\tau$.

The weighting estimator, while unbiased, is generally not efficient as it does not leverage additional information from the outcome model \cite{robins2000marginal}. To further enhance efficiency, we can estimate the conditional expectation of the outcome using a generalized linear model, denoted as $\mu_a(\bm{X};\bm{\theta_a})$, and calculate the residuals as
$$
R(S, A, Y; \bm\theta_1, \bm\theta_0) = S\{Y - A\mu_1(\bm{X}; \bm\theta_1) - (1-A)\mu_0(\bm{X}; \bm\theta_0)\}.
$$
The estimated residual is given by
$$
\hat{R}_i = R(S_i, A_i, Y_i; \hat{\bm\theta}_1, \hat{\bm\theta}_0),
$$
allowing for the use of an augmented weighted estimator:
$$\hat\tau_{aw} = \frac{\sum_{i=1}^n \hat w_{1i}\hat R_i}{\sum_{i=1}^n \hat w_{1i}} - \frac{\sum_{i=1}^n \hat w_{0i}\hat R_i}{\sum_{i=1}^n \hat w_{0i}}  +  \frac{1}{n_2}\sum_{i=1}^n(1-S_i)\left\{\mu_1(\bm{X}_i;\hat{\bm\theta}_1) - \mu_0
(\bm{X}_i;\hat{\bm \theta}_0)\right\}$$

The augmented weighting estimator is doubly robust, that is when either \( h_s(\bm X; \bm\beta) \) and \( e_a(\bm X; \bm\alpha) \) are correctly specified, or \( \mu(a, \bm X;\bm \theta_a) \) is correctly specified, \(\hat{\tau}_{aw}\) is consistent for \(\tau\). 

Furthermore, under the aforementioned assumptions, the asymptotic variance of estimator for $\tau$ follows the semiparametric efficiency bound \cite{tsiatis2006semiparametric} which is the smallest asymptotical variance in any semiparametic model. Both the weighting and augmented weighting estimators follow this bound, as they rely on the empirical distribution of $Y$. We establish this bound in Theorem 1 for the non-nested setting, just as \citet{li2023note} established it for the nested setting.

\begin{theorem} \label{theorem1}
Under the aforementioned assumptions, the semiparametric efficiency bound for any regular asymptotical linear semiparametric estimator of $\tau$ is
$$V(\tau) = \frac{1}{(1-h_s)^2}\mathbb{E}\left[\left\{\frac{(1-h_s(\bm{X}))^2}{h_s(\bm{X})}\right\}\left(\frac{\sigma_1^2(\bm{X})}{e_1(\bm{X})} + \frac{\sigma_0^2(\bm{X})}{e_0(\bm{X})}\right) + (\tau(\bm{X}) - \tau)^2\right]$$

where $h_s$ denotes $Pr(S=1|D=1)$, $\sigma_1^2(\bm{X})$ denotes $Var(Y(1)|\bm{X}, A = 1)$, $\sigma_0^2(\bm{X})$ denotes $Var(Y(0)|\bm{X})$, and $\tau(\bm{X}) = \mathbb{E}(Y(1)-Y(0)|\bm{X})$.

\end{theorem}

Theorem 1 highlights that design factors, such as sampling scores and propensity scores, play a crucial role in determining the semiparametric efficiency bounds. If, for some $\bm{x}$ with a positive density function $f_{X|S=0}(\bm{x})$, $h_s(\bm{x})e_a(\bm{x}) \to 0$, then $V(\tau) \to \infty$. This implies that challenges at the design stage, such as highly biased sampling or treatment assignment, can result in a large variance bound for any semiparametric estimator. Consequently, this underscores the importance of the positivity assumption and sufficient overlap in ensuring reliable estimation.

\section{Proposed Methods for Addressing Positivity Violation}
As indicated by Theorem 1, an extremely small product of the propensity scores and sampling scores, which can arise from issues at the design stage, lead to unreliable estimation. In fact, the estimator may not even exist if either of these scores are equal to zero. One common strategy to address this is to shift the focus of estimation to a transportable population\cite{huang2024overlap}, which serves as a bridge between the target population and the observed sample. A good transportable population has an acceptable variance in the estimation of its ATE and closely resembles the target population. A widely accepted approach identifies this transportable population as a subset of the target population by excluding subjects with extreme propensity score and sampling score \cite{chen2023generalizability, parikh2024we}. However, it is important to note that the ATE for the transportable population can differ from that of the original target population.

To address positivity violations when inferring the ATE of the original target population, it is essential to go beyond merely considering the transportable population. Therefore, we propose a novel framework to estimate the ATE of the original target population under conditions of near or full positivity violations. This framework consists of three main steps: (1) dividing the target population into unrepresented, underrepresented, and well-represented groups based on the observed samples; (2) using weighting estimators to estimate the ATE for the well-represented group; and (3) conducting a sensitivity analysis to infer the ATE for the original target population.

\subsection{Division of The Target Population}
\begin{figure}[!h]
    \centering
    \includegraphics[width=0.8\textwidth]{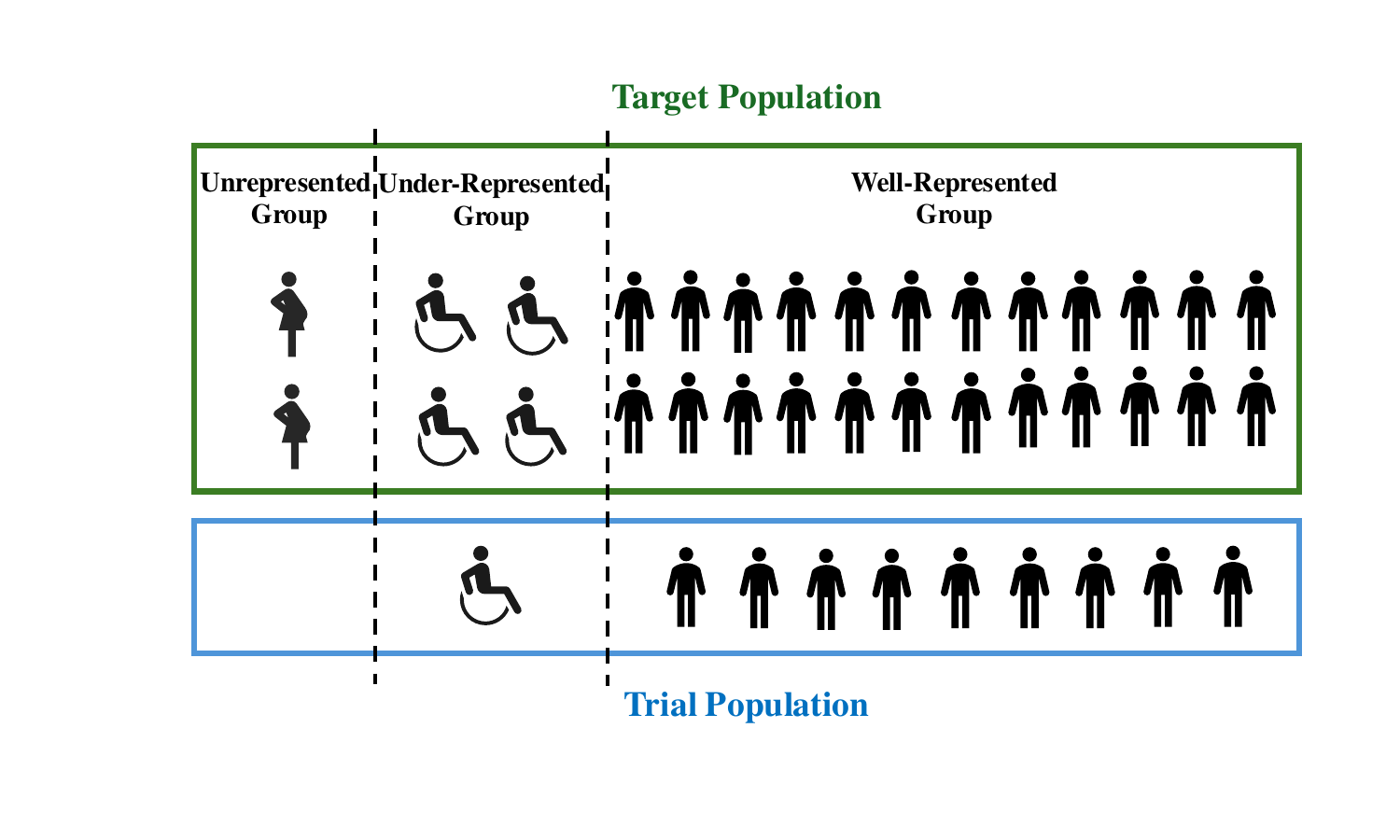}
    \caption{Demonstration of the division of the target population based on patients' characteristics.}
    \label{fig:fig2}
\end{figure}

In the first step, we divide individuals within the target population into unrepresented, underrepresented, and well-represented groups, as illustrated in Figure \ref{fig:fig2}. This division is based on the extent to which each group is represented in the observed sample, providing a foundation for subsequent weighting and sensitivity analysis. The three groups are defined as follows:

\begin{itemize}
    \item \textbf{Unrepresented Group} \(\mathcal{R}_1\): This group includes individuals for whom no comparable subjects were observed in the study sample, either because their probability of participation in the study is zero, \(h_s(\bm{X}) = 0\), or because they lack comparable subjects in the opposite treatment group, that is, \(e_a(\bm{X}) = 0\) or \(1 - e_a(\bm{X}) = 0\). Formally, this group is defined as \(\mathcal{R}_1 = \{ \bm{X} \in \mathcal{X}_{\text{excl}} \}\), where \(\mathcal{X}_{\text{excl}}\) represents the region of the covariate space for these subjects, determined by the known study inclusion criteria and treatment assignment rules.

    \item \textbf{Underrepresented Group} \(\mathcal{R}_2\): This group includes individuals for whom the sampling score or the propensity score for one treatment is very small. It is defined as \(\mathcal{R}_2 = \{ \bm{X} \mid h_s(\bm{X})e_a(\bm{X}) < \delta \text{ and } h_s(\bm{X})(1 - e_a(\bm{X})) < \delta, \text{ and } \bm{X} \notin \mathcal{X}_{\text{excl}} \}\), where \(\delta\) is a small threshold value indicating extreme underrepresentation.

    \item \textbf{Well-Represented Group} \(\mathcal{R}_3\): This group includes individuals for whom the sampling score and propensity score for both treatments are not extremely small, ensuring adequate representation. It is defined as \(\mathcal{R}_3 = \{ \bm{X} \mid \bm{X} \notin \mathcal{R}_1 \cup \mathcal{R}_2 \}\).
\end{itemize}

Accordingly, the ATE for the target population can be expressed as a weighted sum of the ATEs for each group:
$$\tau = p_1\tau_1 + p_2\tau_2 + p_3\tau_3$$
where \(\tau_j = \mathbb{E}(Y(1) - Y(0) \mid \bm{X} \in \mathcal{R}_j, S = 0)\), and \(p_j\) for \(j = 1, 2, 3\) represents the proportion of each corresponding group in the target population. Additionally, \(\tau_{12} = \mathbb{E}(Y(1) - Y(0) \mid \bm{X} \in \mathcal{R}_1 \cup \mathcal{R}_2, S = 0)\).

In this division, the groups \(\mathcal{R}_1\) and \(\mathcal{R}_2\) need to be separated because the sampling or propensity scores for individuals in \(\mathcal{R}_1\) are exactly zero due to the study’s inclusion criteria and strict treatment assignment rules. This means that their scores do not require estimation, and their ATE is unidentifiable. Additionally, imputing outcomes for \(\mathcal{R}_1\) requires extra assumptions through statistical modeling, as outcomes for individuals in \(\mathcal{R}_1\) are not observed. This challenge is particularly relevant when the rules defining \(\mathcal{R}_1\) involve categorical variables \cite{zivich2023synthesis}. Furthermore, \(\mathcal{R}_2\) and \(\mathcal{R}_3\) must also be separated because subjects in \(\mathcal{R}_2\) disproportionately inflate the variance in ATE estimation, as indicated by Theorem~\ref{theorem1}. This separation allows us to identify a subset of the target population—the well-represented group—that can be efficiently estimated while also enabling investigators to better report the limitations of their study sample.

However, determining an appropriate threshold \( \delta \) to separate \( \mathcal{R}_2 \) and \( \mathcal{R}_3 \) is often challenging. To address this, we propose defining \( \delta \) based on the desired proportion of subjects to retain in the well-represented group for inference, similar to trimming based on quantiles \cite{cole2008constructing, sturmer2010treatment, lee2011weight}. This approach offers several advantages: it allows flexibility in controlling the proportion of data retained, helping to balance estimation efficiency and robustness. It is adaptable to various data distributions and does not rely on distributional assumptions, making it a practical choice when external information is limited. Finally, the concept of being “well-represented” is inherently relative, as the over-representation of some groups leads to the under-representation of others, making a proportion or quantile-based approach a logical choice to account for these imbalances.

To facilitate later estimation, the membership in \( \mathcal{R}_3 \) is defined by the inclusion weight
\[
\begin{array}{c}
\mathcal{R}_3 = \{\bm{X} \mid k(\bm{X}; \delta) = 1 \},\\

k(\bm{X}; \delta) = I(h_s(\bm{X})e_1(\bm{X}) \ge \delta) I((e_0(\bm{X}))h_s(\bm{X}) \ge \delta).
\end{array}
\]
If the desired proportion of well-represented subjects is \( p_3^* \), the threshold \( \delta^* \) is chosen to satisfy the following condition:
$$\mathbb{E}(k(\bm{X}; \delta^*) \mid S = 0) = p_3^*.$$
Since $\delta^*$ must be estimated from the data, we suggest employing a smooth inclusion function, as proposed by \citet{yang2018asymptotic}, which allows for conventional asymptotic linearization techniques to analyze the resulting estimators. Specifically, we define the smooth inclusion function as
$$k_s(\bm X; \delta, \epsilon, \bm \alpha, \bm \beta) = \Phi_{\epsilon, \delta}(h_s(\bm{X};\bm\beta)e_1(\bm{X};\bm\alpha))\Phi_{\epsilon, \delta}(h_s(\bm{X};\bm\beta)e_0(\bm{X};\bm{\alpha})),$$
where \(\Phi_{\epsilon, \delta}(\cdot)\) represents the cumulative distribution function of a normal distribution with mean \(\delta\) and a fixed variance \(\epsilon^2\) (e.g. \(\epsilon = 10^{-8}\)). The smooth function \(k_s(\bm{X}; \delta, \epsilon)\) approaches the original indicator-based function \(k(\bm{X}; \delta)\) as \(\epsilon \to 0\). 

Accordingly, the smooth inclusion weight for observed subjects, based on the estimated propensity score and sampling score, is given by
$$\hat k_{si}(\delta) = \Phi_{\epsilon, \delta}(h_s(\bm{X}; \hat{\bm\beta})e_1(\bm{X}; \hat{\bm\alpha})) \Phi_{\epsilon, \delta}(h_s(\bm{X}; \hat{\bm\beta}) e_0(\bm{X}; \hat{\bm\alpha})).$$
%


\subsection{Inference in Well-Represented Group}
After dividing the target population into three groups, the inference for the well-represented group can be obtained using the aforementioned Hájek weighting estimator or the augmented weighting estimator with additional defined inclusion weights.

Specifically, the required parameter \( \bm \Xi_w = [\bm\beta, \bm\alpha, \delta^*, \mu_{31}, \mu_{30}]^T \) for the weighting estimator is obtained by solving the estimating equation
\[
\sum_{i=1}^n \bm \Psi_w(S_i, \bm{X}_i, A_i, Y_i; \hat{\bm \Xi}_w) = 0,
\]
where the estimating function \( \bm\Psi_w(S, \bm{X}, A, Y; \bm \Xi_w) \) is defined as
\[
\bm\Psi_w(S, \bm{X}, A, Y; \bm \Xi_w) = 
\begin{bmatrix}
\bm \Psi_{\bm\beta}\\
\bm \Psi_{\bm\alpha}\\
\Psi_{\delta^*} \\
\Psi_{\mu_{30}}\\
\Psi_{\mu_{31}}\\
\end{bmatrix} = \begin{bmatrix}
\{S - h_1(\bm{X}^T\bm\beta)\}\bm{X} \\
S\{A - h_1(\bm{X}^T\bm\alpha)\}\bm{X} \\
(1 - S) k_s(\bm{X}; \delta, \epsilon, \bm\alpha, \bm\beta) - p_3^* \\
k_s(\bm{X}; \delta, \epsilon, \bm\alpha, \bm\beta)w_{1}(S, A, \bm{X}; \bm\alpha, \bm\beta)(Y - \mu_{31})\\
k_s(\bm{X}; \delta, \epsilon, \bm\alpha, \bm\beta)w_{0}(S, A, \bm{X}; \bm\alpha, \bm\beta)(Y - \mu_{30})\\
\end{bmatrix}.
\]
where \( \bm \Psi_{\bm\beta} \) and \( \bm \Psi_{\bm\alpha} \) are the score function vectors for the sampling score and propensity score models, respectively, with \( h_1(\cdot) \) as the link function. \( \Psi_{\mu_{30}} \) and \( \Psi_{\mu_{31}} \) represent the estimating functions for the mean outcomes in the control and treatment groups within the well-represented group. And the weighting estimator for well-represented group \( \hat{\tau}_{3w} \) is given by
\begin{equation}
\label{eq:1}
\hat{\tau}_{3w} = \hat{\mu}_{31} - \hat{\mu}_{30}.
\end{equation}
Similarly, the required parameter  \( \bm\Psi_{aw}(S, \bm X, A, Y;\bm \Xi_{aw}) \) is obtained by the estimating equation
$$\sum_{i=1}^n \bm \Psi_{aw}(S_i, \bm X, A, Y; ;\hat{\bm \Xi}_{aw}) = 0 $$
where the estimating function \( \bm\Psi_w(S, \bm{X}, A, Y; \bm \Xi_w) \) is defined as 
$$
\bm\Psi_{aw}(S, \bm{X}, A, Y;\bm \Xi_{aw}) = 
\begin{bmatrix}
\bm \Psi_{\bm\beta}\\
\bm \Psi_{\bm\alpha}\\
\Psi_{\bm\theta_1} \\
\Psi_{\bm\theta_0} \\
\Psi_{\delta^*} \\
\Psi_{v_1}\\
\Psi_{v_2}\\
\Psi_{v_3}
\end{bmatrix} = \begin{bmatrix}
\{S - h_1(\bm{X}^T\bm\beta)\}\bm{X} \\
S\{A - h_1(\bm{X}^T\bm\alpha)\}\bm{X} \\
SA_i\{Y_i - h_2(\bm X^T\bm\theta_1)\}\bm X\\
S(1-A_i)\{Y_i - h_2(\bm X^T\bm\theta_0)\}\bm X\\
(1 - S) k_s(\bm X; \delta, \epsilon, \bm \alpha, \bm \beta) - p_3^* \\
k_s(\bm X; \delta, \epsilon, \bm \alpha, \bm \beta)w_{1}(S, A, \bm X;\bm \alpha, \bm \beta)\{R(S, A, Y; \bm\theta_0, \bm\theta_1) - v_1\}\\
k_s(\bm X; \delta, \epsilon, \bm \alpha, \bm \beta)w_{0}(S, A, \bm X;\bm \alpha, \bm \beta)\{R(S, A, Y; \bm\theta_0, \bm\theta_1) - v_2\}\\
k_s(\bm X; \delta, \epsilon, \bm \alpha, \bm \beta)(1-S)\{\mu_1(\bm X; \bm \theta_1) - \mu_0(\bm X; \bm \theta_0)- v_3\}\\
\end{bmatrix}
$$
where \( \bm \Psi_{\bm\theta_1} \) and \( \bm \Psi_{\bm\theta_0} \) are the score function vectors for the outcome regression in the treatment and control groups, respectively, and \( h_2(\cdot) \) is the link function for their generalized models. \( \bm \Psi_{v_1} \) and \( \bm \Psi_{v_2} \) are the estimating functions for the weighted average residuals in the treatment and control groups. \( \bm \Psi_{v_3} \) represents the estimating function for the ATE based on the outcome regression. And the augmented weighting estimator for well-represented group \( \hat{\tau}_{3aw} \) is given by
\begin{equation}
    \label{eq:2}
\hat\tau_{3aw} = \hat v_1 - \hat v_2 + \hat v_3.
\end{equation}
For the two proposed estimators, the following Theorem \ref{thm:2} and Theorem \ref{thm:3} are established using the M-estimation theory and the delta method to support their inference.

\begin{theorem}
\label{thm:2}
The weighting estimator
\(\hat{\tau}_{3w}\) in (\ref{eq:1}) is asymptotically linear. Furthermore,
$$
\sqrt{n}(\hat{\tau}_{3w} - \tau_{3}) \rightarrow N(0, \bm{\eta}_w^T \bm{A}_w^{-1} \bm{B}_w \bm{A}_w^{-T} \bm{\eta}_w)
$$
where 
\begin{align*}
\bm{A}_w &= \mathbb{E}\left(\frac{\partial}{\partial \bm{\Xi}_w^T} \Psi_w(S_i, \bm{X}_i, A_i, Y_i)\right), \\
\bm{B}_w &= \mathbb{E}\left(\Psi_w(S_i, \bm{X}_i, A_i, Y_i) \Psi_w^T(S_i, \bm{X}_i, A_i, Y_i)\right), \\
\bm{\eta}_w &= [\bm 0_{1 \times p}, \bm 0_{1 \times p}, 0, 1, -1].
\end{align*}
\end{theorem}

\begin{theorem}
\label{thm:3}
The augmented weighting estimator 
\(\hat{\tau}_{3aw}\) in (\ref{eq:2}) is asymptotically linear. Furthermore,
$$
\sqrt{n}(\hat{\tau}_{3aw} - \tau_{3}) \rightarrow N(0, \bm{\eta}_{aw}^T \bm{A}_{aw}^{-1} \bm{B}_{aw} \bm{A}_{aw}^{-T} \bm{\eta}_{aw}),
$$
where 
\begin{align*}
\bm{A}_{aw} &= \mathbb{E}\left(\frac{\partial}{\partial \bm{\Xi}_{aw}^T} \Psi_{aw}(S_i, \bm{X}_i, A_i, Y_i)\right), \\
\bm{B}_{aw} &= \mathbb{E}\left(\Psi_{aw}(S_i, \bm{X}_i, A_i, Y_i) \Psi_{aw}^T(S_i, \bm{X}_i, A_i, Y_i)\right), \\
\bm{\eta}_{aw} &= [\bm 0_{1 \times p}, \bm 0_{1 \times p}, \bm 0_{1 \times p}, \bm 0_{1 \times p}, 0, 1, -1, 1].
\end{align*}
\end{theorem}

The above theorems indicate that the variance for the proposed weighting estimator based on the smooth inclusion weights under the estimated threshold can be estimated using either bootstrapping or a sandwich variance estimator.

\subsection{Inference in Target Population}
Building on the estimation in the well-represented group, we now extend the inference to the target population. Since direct inference from observed data for unrepresented and underrepresented groups is either infeasible or highly variable, we propose two approaches to estimate the ATE for these groups using sensitivity parameters.

In the first, \textbf{Group Proportional Difference} approach, we infer the ATE for unrepresented and underrepresented groups by assuming a proportional difference relative to the well-represented group. Specifically,
    the ATE for the unrepresented and underrepresented groups is assumed to be \( k_1 \)  and \( k_2 \) times the ATE of the well-represented group.
Thus, the estimated total ATE is
$$\hat{\tau} = (k_1\hat p_1 + k_2\hat{p}_2 + p_3^*)\hat\tau_3$$



When \(k_1 = 
 k_2 = 1\), we assume that there is no difference between the ATE of well-represented groups and the ATE of unrepresented and underrepresented groups.  Therefore, the ATE calculated from the well-represented group is taken as the ATE for the entire target population. This approach aligns with scientific intuition: if these groups constitute a larger portion of the population or display greater treatment effect differences, their impact on the overall sensitivity analysis will naturally be more pronounced.

The second approach relies on the extrapolation of $\tau_1$ and $\tau_2$ using either a statistical or mathematical model.
Let
$$ \begin{array}{c}
\zeta_{1} = \frac{1}{n_1} \sum_{i: \bm{X}_i \in \mathcal{R}_1} \left( m_1(1, \bm{X}_i) - m_1(0, \bm{X}_i) \right)\\
\zeta_{2} = \frac{1}{n_2} \sum_{i: \bm{X}_i \in \mathcal{R}_2} \left( m_2(1, \bm{X}_i) - m_2(0, \bm{X}_i) \right)
\end{array}
$$
where $\zeta_1$ and $\zeta_2$ are the group ATEs under model extrapolation for the unrepresented and underrepresented groups,  and \( m_1(a, \bm{X}_i) \) and \( m_2(a, \bm{X}_i) \) are the extrapolation models. Various extrapolation models have been proposed in the literature, including outcome models \cite{nethery2019estimating, zhu2023addressing}, propensity score regression models \cite{ma2020robust}, and investigator-specified mathematical models \cite{zivich2023synthesis, zivich2024transportability}.


We further make the
%
\textbf{Extrapolation Proportional Difference} assumption, that the ATE for the unrepresented and underrepresented groups is assumed to be \( k_1 \) and \( k_2 \) times their model-based extrapolation results

The overall estimated ATE is then given by
$$
\hat{\tau} =  \hat{p}_1k_1\zeta_1 + \hat{p}_2k_2\zeta_2 + p_3^*\hat{\tau}_3. 
$$
%

The sensitivity parameter \( k \) is introduced for extrapolation results because both the point estimate and uncertainty quantification are highly influenced by model specifications for underrepresented or unrepresented groups. 
Under this sensitivity approach, the impact of underrepresented and unrepresented groups depends on three factors: the proportion of subjects in these groups, the extrapolation result, and the proportional difference between the extrapolation result and true ATE of unrepresented and underrepresented groups. 

Following these three steps enables us to assess the trial sampling mechanism across the three groups, make the inference in the well-represented group, and examine the robustness of the findings for the entire target population through sensitivity analysis.


\section{Simulation Study}
In this section, we present simulation studies that assess the finite-sample performance of the proposed methods for continuous outcomes. Simulation settings and results for binary outcomes, as well as the corresponding R code, are provided in the Supplementary materials. For inference, we assume the sensitivity parameter to be known. Specifically, in the first approach, we assume that the proportional differences between the groups were known, while in the second approach, we assume that the ATE for the unrepresented or underrepresented groups is known. Based on these assumptions, we evaluated the mean bias, mean squared error, standard deviation, and coverage rate of the proposed estimators over 1,000 replications.

\subsection{Data Generation}
We simulate cohorts with total sample sizes of $\tilde{n} = 20000$, $50000$, and $100000$ individuals, with corresponding study sample sizes $n_1$ of approximately 200, 500, and 1000. Specifically, we generate baseline covariates $\bm{X_i} = (1, X_{1i}, X_{2i}, X_{3i}, X_{4i})$, where $X_{ji} \sim N(0, 1)$ for $j = 1, 2, 3, 4$. In addition, we include a binary covariate $E \sim Ber(0.01)$. Samples are strictly excluded from the study when $E = 1$ or $X_4 \ge 3$. The selection in the study is simulated using a binary indicator $S_i$, where $\Pr[S=1|\bm{X}, E=0, X_4<3] = \frac{1}{1+\exp(-\bm{X}^\top \bm{\beta})}$ with $\bm{\beta} = [-7.523499, -2, 1, 1, 1]$, and $\Pr[S=1|E=1 \text{ or } X_4 \ge 3] = 0$. We choose intercept of in $\bm{\beta}$ using numerical methods proposed by \citet{austin2008performance}, such that the study participation probability is 1\%. We then randomly sample 10\% of the non-participants as observed samples with baseline covariates for nonparticipants. The sample size of selected non-participants $n_2$ is approximately 1980, 4950, and 9990. Potential outcomes are generated as $Y(a) \sim N (\bm{X}^\top \bm{\theta}_a - 0.5aE)$, with $\bm{\theta}_0 = [1, 2, 2, 1, 1]$ and $\bm{\theta}_1 = [0, 1, 1, 1, 1]$. For study participants, we generated treatment assignment $A$, where $A \sim \text{Bernoulli}(0.5)$. The observed outcome for study participants ($S_i = 1$) is $Y_i = AY_i(1) + (1-A)Y_i(0)$. 

For the simulated dataset across three sample size settings, we estimate the well-represented group proportions at $90\%$ and $80\%$ using both the weighting and augmented weighting estimators. Subsequently, inference for the original target population is conducted with known sensitivity parameters using each of the two approaches. This setup resulted in a total of 8 estimators for each sample size setting. 

\subsection{Simulation Result}
Tables \ref{tab:table1} present simulation results for eight estimators across three different sample size settings. All estimators are unbiased and consistently converge to the true value as the sample size increases, with coverage rates reaching nominal levels in larger samples. Among the estimators, the augmented weighting estimator with correct model specification consistently demonstrated lower MSE compared to the standard weighting estimator. Efficiency gains were observed in both estimators when the proportion of well-represented groups was 0.8, compared to 0.9 in our simulation setting. However, the optimal proportion for efficiency gains varies based on the data, making its determination challenging in real-world settings. However, these efficiency gains come at the cost of a lower proportion of the well-represented group, amplifying the impact of the sensitivity parameters when they are unknown.




\newpage

\begin{table}[H]
\caption{Simulation Results. This table presents simulation results for various trial and target sizes, comparing two methods (IPW and AIPW) under two assumptions (EPD and GPD) and across different proportions of well-represented groups.}
\label{tab:table1}
\centering
\begin{threeparttable}
\begin{tabular}{|r|r|r|l|l|r|r|r|r|}
\hline
\textbf{Trial Size} & \textbf{Target Size} & \textbf{Proportion} & \textbf{Method} & \textbf{Assumption} & \textbf{Bias} & \textbf{MSE} & \textbf{SD} & \textbf{Coverage} \\
\hline
200 & 1980 & 0.8 & IPW & EPD & 0.089 & 1.187 & 1.087 & 0.909 \\
200 & 1980 & 0.8 & IPW & GPD & 0.134 & 2.414 & 1.549 & 0.908 \\
500 & 4950 & 0.8 & IPW & EPD & 0.059 & 0.700 & 0.835 & 0.925 \\
500 & 4950 & 0.8 & IPW & GPD & 0.092 & 1.424 & 1.190 & 0.922 \\
1000 & 9990 & 0.8 & IPW & EPD & -0.007 & 0.504 & 0.710 & 0.923 \\
1000 & 9990 & 0.8 & IPW & GPD & 0.001 & 1.241 & 1.120 & 0.926 \\
200 & 1980 & 0.8 & AIPW & EPD & -0.011 & 0.097 & 0.313 & 0.920 \\
200 & 1980 & 0.8 & AIPW & GPD & -0.015 & 0.200 & 0.449 & 0.920 \\
500 & 4950 & 0.8 & AIPW & EPD & 0.007 & 0.041 & 0.203 & 0.930 \\
500 & 4950 & 0.8 & AIPW & GPD & 0.010 & 0.084 & 0.292 & 0.930 \\
1000 & 9990 & 0.8 & AIPW & EPD & -0.021 & 0.016 & 0.125 & 0.937 \\
1000 & 9990 & 0.8 & AIPW & GPD & -0.030 & 0.033 & 0.180 & 0.937 \\
200 & 1980 & 0.9 & IPW & EPD & 0.034 & 1.399 & 1.188 & 0.940 \\
200 & 1980 & 0.9 & IPW & GPD & 0.041 & 2.001 & 1.421 & 0.940 \\
500 & 4950 & 0.9 & IPW & EPD & 0.118 & 0.974 & 0.985 & 0.930 \\
500 & 4950 & 0.9 & IPW & GPD & 0.141 & 1.394 & 1.178 & 0.930 \\
1000 & 9990 & 0.9 & IPW & EPD & 0.011 & 1.036 & 1.023 & 0.895 \\
1000 & 9990 & 0.9 & IPW & GPD & 0.013 & 1.483 & 1.224 & 0.895 \\
200 & 1980 & 0.9 & AIPW & EPD & -0.013 & 0.140 & 0.376 & 0.920 \\
200 & 1980 & 0.9 & AIPW & GPD & -0.015 & 0.201 & 0.450 & 0.920 \\
500 & 4950 & 0.9 & AIPW & EPD & 0.008 & 0.060 & 0.245 & 0.920 \\
500 & 4950 & 0.9 & AIPW & GPD & 0.010 & 0.085 & 0.293 & 0.920 \\
1000 & 9990 & 0.9 & AIPW & EPD & -0.025 & 0.023 & 0.151 & 0.937 \\
1000 & 9990 & 0.9 & AIPW & GPD & -0.030 & 0.033 & 0.181 & 0.937 \\
\hline
\end{tabular}
\begin{tablenotes}
\footnotesize
\item \textbf{Abbreviations:} IPW, Inverse Probability Weighting; AIPW, Augmented IPW; EPD, Extrapolation Proportional Difference; GPD, Group Proportional Difference.
\end{tablenotes}
\end{threeparttable}
\end{table}

\section{Application}
Opioid use disorder is a serious condition that increases the risk of overdose, relapse, and significant physical, mental, and social harm if left untreated. Medication-assisted treatments, such as buprenorphine (BUP) and methadone (MET), are among the most effective options for managing this disorder. However, treatment success often depends on patient adherence to the prescribed regimen, with nonadherence leading to suboptimal outcomes. To assess the effectiveness of these treatments, the Starting Treatment with Agonist Replacement Therapy (START) trial, a multi-center study, randomized 1,271 participants in a 2:1 ratio to receive either BUP or MET. A secondary analysis of the trial data indicated that MET had significantly higher treatment completion rates than BUP, with 74\% of MET participants completing treatment compared to 46\% of BUP participants (p < 0.01) \cite{hser2014treatment}.

To generalize these findings to a broader real-world population, we used data from the 2021 Treatment Episode Data Set – Admissions (TEDS-A), which includes 166,270 individuals receiving medication-assisted treatment for opioid dependence or abuse. Table \ref{tab:table2} compares the baseline covariates of the START trial sample with those of the TEDS-A population. Our primary goal was to estimate the average treatment effect of BUP versus MET on treatment completion rates in the real-world population. However, positivity violations were observed due to age and pregnancy exclusions in the trial sample, as well as baseline covariates that influenced trial participation. 

\begin{figure}[!h]
    \centering
    \includegraphics[width=0.75\textwidth]{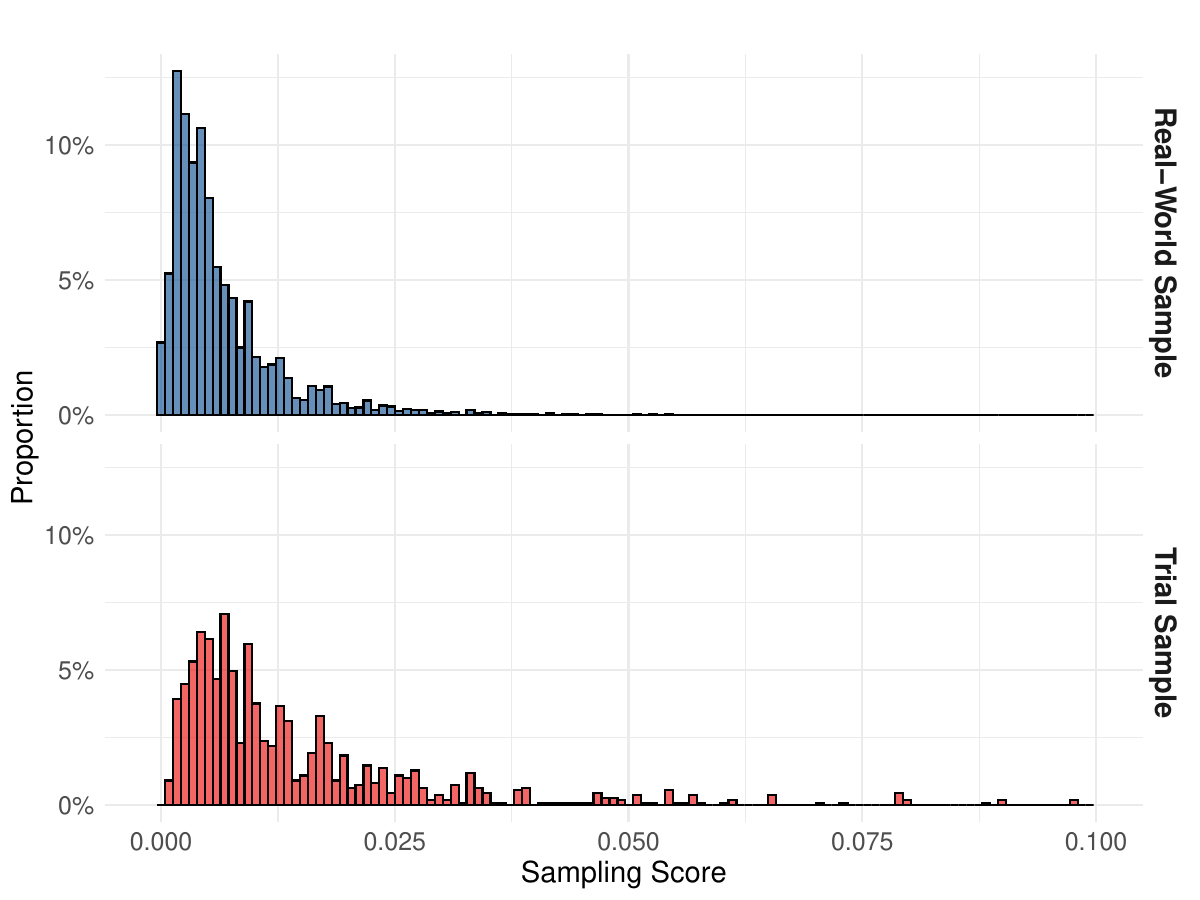}
    \caption{Histogram of sampling score in STAR trial and 2021 TEA data}
    \label{fig:fig3}
\end{figure}

\begin{figure}[h!]
    \centering
    \includegraphics[width=0.65\textwidth]{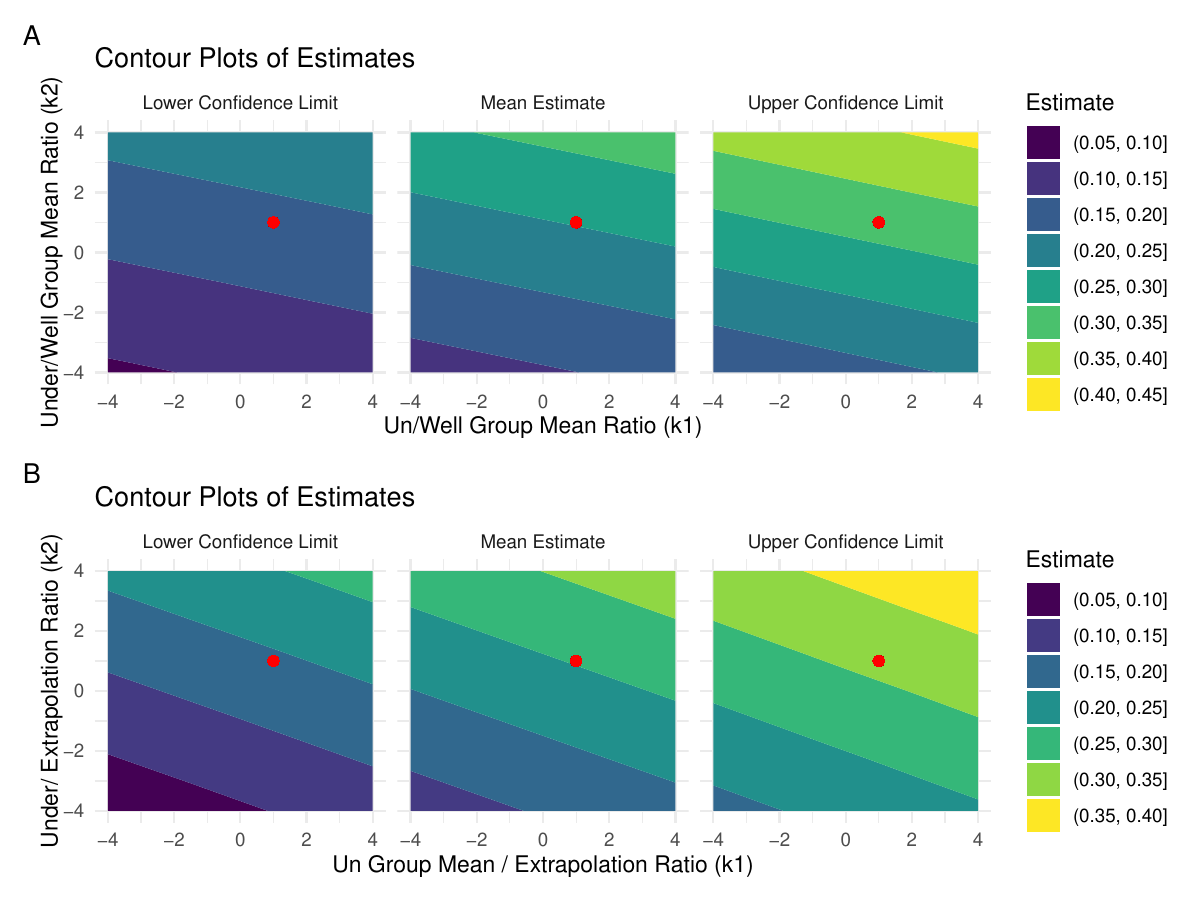}
    \caption{Sensitivity analyses for the average treatment effect on completion rates: (A) Under the Group Proportional Difference Assumption; (B) Under Extrapolation Proportional Difference Assumption. The red points in the figures correspond to sensitivity parameters all set to 1.}
    \label{fig:fig4}
\end{figure}

A logistic regression model, including covariates such as age, sex, race, injection behavior, and other substance use, was built to assess the representativeness of trial recruitment, while treatment assignment remained unbiased due to randomization. Figure \ref{fig:fig3} presents the distribution of estimated probabilities of trial participation for the START trial participants and the TEDS-A population. Compared to the START trial participants, a larger proportion of individuals in the TEDS-A population had extremely low or zero probabilities of participating in the trial. Using the estimated sampling scores, the target TEDS-A population was divided into three groups based on the trial exclusion criteria, with the well-represented group defined as comprising 90\% of the total target population. For the well-represented group in the target population, the ATE is 24.9\% (95\% CI [18.3\%–31.1\%]) based on the augmented weighting estimator. Two sensitivity analysis approaches were applied to address the original question, varying the sensitivity parameter within a range of -4 to 4. Under the GPD assumption, even when the ATE for the underrepresented group and unrepresented group is -4 times that of the well-represented group, the average treatment effect remains significantly greater than 0 for the entire target population (\ref{fig:fig4}). This suggests that if the treatment was applied to the whole target population, Methadone would still result in a higher completion rate. Under the EPD assumption, we use the 18–20 age group to extrapolate for individuals below 18 and the non-pregnant group to extrapolate for pregnant individuals. Two separate logistic models—one for the Methadone group and one for the Buprenorphine/Naloxone group—were built to extrapolate the outcome. The ATE is 38.7\% for the unrepresented group and 22.5\% for the underrepresented group. Based on \ref{fig:fig4}, the study findings remain robust in the original population even under extreme sensitivity parameters.

\begin{table}[H]

\caption{Baseline Characteristics of Study Sample and Target Population. This table presents baseline characteristics comparing the Buprenorphine/Naloxone group, the Methadone group in the trial, and the Target Population.}
\label{tab:table2}
\centering
\begin{threeparttable}
\begin{tabular}{|l|c|c|c|}
\hline
\textbf{Characteristic} & \textbf{Buprenorphine/Naloxone} & \textbf{Methadone} & \textbf{Target Population} \\ 
\hline
\textbf{Sample Size} & 654 & 440 & 166,270 \\ 
\hline
\textbf{Age, years} & & & \\ 
\hline
12-14 & 0 (0.0\%) & 0 (0.0\%) & 4 (0.0\%) \\ 
\hline
15-17 & 0 (0.0\%) & 0 (0.0\%) & 88 (0.1\%) \\ 
\hline
18-20 & 14 (2.1\%) & 2 (0.5\%) & 1,827 (1.1\%) \\ 
\hline
21-24 & 76 (11.6\%) & 52 (11.8\%) & 7,873 (4.7\%) \\ 
\hline
25-29 & 122 (18.7\%) & 88 (20.0\%) & 25,499 (15.3\%) \\ 
\hline
30-34 & 96 (14.7\%) & 67 (15.2\%) & 36,147 (21.7\%) \\ 
\hline
35-39 & 71 (10.9\%) & 40 (9.1\%) & 29,558 (17.8\%) \\ 
\hline
40-44 & 69 (10.6\%) & 58 (13.2\%) & 20,724 (12.5\%) \\ 
\hline
45-49 & 88 (13.5\%) & 57 (13.0\%) & 12,785 (7.7\%) \\ 
\hline
50-54 & 66 (10.1\%) & 49 (11.1\%) & 11,768 (7.1\%) \\ 
\hline
55-64 & 50 (7.6\%) & 25 (5.7\%) & 16,063 (9.7\%) \\ 
\hline
65+ & 2 (0.3\%) & 2 (0.5\%) & 3,934 (2.4\%) \\ 
\hline
\textbf{Male} & 451 (69.0\%) & 300 (68.2\%) & 105,980 (63.7\%) \\ 
\hline
\textbf{Pregnancy} & & & \\ 
\hline
Yes & 0 (0.0\%) & 0 (0.0\%) & 2,980 (1.8\%) \\ 
\hline
No & 203 (31.0\%) & 140 (31.8\%) & 57,310 (34.5\%) \\ 
\hline
Not applicable & 451 (69.0\%) & 300 (68.2\%) & 105,980 (63.7\%) \\ 
\hline
\textbf{Race} & & & \\ 
\hline
White & 433 (66.2\%) & 308 (70.0\%) & 108,176 (65.1\%) \\ 
\hline
Hispanic & 107 (16.4\%) & 63 (14.3\%) & 28,609 (17.2\%) \\ 
\hline
Black & 59 (9.0\%) & 40 (9.1\%) & 20,385 (12.3\%) \\ 
\hline
Other & 55 (8.4\%) & 29 (6.6\%) & 9,100 (5.5\%) \\ 
\hline
\textbf{Substance Use} & & & \\ 
\hline
Injection drug use & 441 (67.4\%) & 306 (69.5\%) & 77,319 (46.5\%) \\ 
\hline
Alcohol use & 141 (21.6\%) & 110 (25.0\%) & 21,343 (12.8\%) \\ 
\hline
Amphetamines use & 70 (10.7\%) & 50 (11.4\%) & 27,934 (16.8\%) \\ 
\hline
Cannabis use & 124 (19.0\%) & 99 (22.5\%) & 24,698 (14.9\%) \\ 
\hline
Sedatives use & 94 (14.4\%) & 59 (13.4\%) & 13,342 (8.0\%) \\ 
\hline
Cocaine use & 206 (31.5\%) & 151 (34.3\%) & 41,174 (24.8\%) \\ 
\hline
\textbf{Completed Study} & 303 (46.3\%) & 325 (73.9\%) & - \\ 
\hline
\end{tabular}
\begin{tablenotes}
\footnotesize
\item \textbf{Abbreviations:} Percentages are shown in parentheses for each characteristic. 
\item \textbf{Note:} "Not applicable" refers to characteristics irrelevant to certain groups (e.g., pregnancy for males).
\end{tablenotes}
\end{threeparttable}
\end{table}

{
\begin{itemize}
    \item \textbf{Unrepresented Group (3,071; 2\%)}: Individuals excluded by trial criteria, including those under 18 years old or pregnant.
    \item \textbf{Underrepresented Group (13,487; 8\%)}: Individuals with extreme weights. This group includes:
    \begin{itemize}
        \item Black individuals aged 55-64 years who do not depend on other drugs and do not use injection drugs (4,456).
        \item Black males aged 18-20, 30-34, or 65 years and older who do not depend on other drugs and do not use injection drugs (1,164).
        \item Other individuals (7,867).
    \end{itemize}
    \item \textbf{Well-represented Group (149,712; 90\%)}
\end{itemize}
}

\section{Disscusion}
In this paper, we propose an integrative framework to address positivity violations when estimating the ATE of a target population. After performing positivity checks and calculating the sampling and propensity scores, the target population is divided into three parts. The well-represented portion is inferred with confidence, while the remaining two parts are inferred with the help of a sensitivity parameter. This approach combines positivity checking, trimming, and sensitivity analysis into a unified framework.

A variety of methods are available for determining the separation threshold for trimming or truncation when addressing violations of the positivity assumption. In addition to quantile-based and fixed threshold criteria, the separation threshold can also be set by minimizing the efficiency bound \cite{chen2023generalizability, parikh2024we, crump2009dealing, khan2024doubly}, proximal mean squared error \cite{xiao2013comparison}, or asymptotic mean squared error \cite{ma2020robust}. However, the asymptotic properties when minimizing the efficiency bound are only guaranteed when the uncertainty in estimating the threshold is ignored. 
Meanwhile, minimizing the proximal mean squared error lacks any established asymptotic theory. \citet{ma2020robust} established the asymptotic behavior of estimation under minimizing the asymptotic mean squared error criterion, and extending this approach to settings involving a target population would be valuable. Our sensitivity analysis remains compatible with group division using these methods. We prefer to incorporate the uncertainty in group division and control the introduced bias by chosen proportion, which is why we choose quantile-based criteria. In practice, researchers may wish to assess the ATE for well-represented groups at different proportions and check the trade-off of trimmed proportion and gained efficiency. One practical approach is to perform simultaneous inference across a set of proportions, as suggested by \citet{khan2024doubly}.

In addition to identifying the transportable population as a subset of the target population, entropy weighting and its modifications have been proposed as alternatives that identify the transportable population using moment constraints to align the mean or other aspects of the covariate distribution \cite{kaizar2023reweighting, chattopadhyay2024one}. However, entropy weighting relies on untestable parametric assumptions about the outcome given the covariates. These additional assumptions also necessitate other corresponding sensitivity analyses to assess their robustness. \citet{leger2022causal} compared several methods under a near-violation of positivity; however, all methods were affected, highlighting the need for additional sensitivity analyses.

In the future, we aim to extend our framework to handle ordinal or continuous treatments, where positivity violations are more likely to occur. Additionally, we plan to integrate semiparametric models into the estimation of propensity scores, sampling scores, and outcome regressions, as these models offer greater flexibility. 

\section{ACKNOWLEDGEMENTS}
The authors declare no conflicts of interest. The population data used in this study are available upon request from the Substance Abuse and Mental Health Services Administration (SAMHSA), and the RCT data are available upon request from the National Institute on Drug Abuse (NIDA) Data Share site.

\newpage

\bibliography{sample}

\end{document}


\clearpage
\appendix
\section*{Appendix}
\addcontentsline{toc}{section}{Appendix}

\section{Simulation for Binary Outcome} 
In this section, we evaluate the simulation performance for a binary outcome. The simulation setup remains the same as in the paper, except that \( y(a) \) is generated from a Bernoulli distribution with probability \( \frac{\exp(\mu_{ai})}{1 + \exp(\mu_{ai})} \), where \( \mu_{ai} = \bm{\theta}_a - 0.5aE \). The simulation result is presented in Supplementary Table 1.

\begin{table}[ht]
\caption{Simulation Results. This table presents simulation results for various trial and target sizes, comparing two methods (IPW and AIPW) under two assumptions (EPD and GPD) and across different proportions of well-represented groups.}
\label{tab:table1}
\centering
\begin{threeparttable}
\begin{tabular}{|r|r|r|l|l|r|r|r|r|}
\hline
\textbf{Trial Size} & \textbf{Target Size} & \textbf{Proportion} & \textbf{Method} & \textbf{Assumption} & \textbf{Bias} & \textbf{MSE} & \textbf{SD} & \textbf{Coverage} \\
\hline
200 & 1980 & 0.8 & IPW & EPD & 0.008 & 0.029 & 0.171 & 0.960 \\
200 & 1980 & 0.8 & IPW & GPD & 0.017 & 0.061 & 0.248 & 0.960 \\
500 & 4950 & 0.8 & IPW & EPD & -0.012 & 0.018 & 0.136 & 0.970 \\
500 & 4950 & 0.8 & IPW & GPD & -0.013 & 0.039 & 0.198 & 0.980 \\
1000 & 9990 & 0.8 & IPW & EPD & -0.004 & 0.013 & 0.113 & 0.930 \\
1000 & 9990 & 0.8 & IPW & GPD & 0.001 & 0.031 & 0.176 & 0.930 \\
200 & 1980 & 0.8 & AIPW & EPD & -0.005 & 0.008 & 0.089 & 0.950 \\
200 & 1980 & 0.8 & AIPW & GPD & -0.009 & 0.016 & 0.126 & 0.950 \\
500 & 4950 & 0.8 & AIPW & EPD & -0.003 & 0.004 & 0.063 & 0.960 \\
500 & 4950 & 0.8 & AIPW & GPD & -0.007 & 0.010 & 0.098 & 0.970 \\
1000 & 9990 & 0.8 & AIPW & EPD & -0.002 & 0.002 & 0.045 & 0.940 \\
1000 & 9990 & 0.8 & AIPW & GPD & -0.005 & 0.005 & 0.071 & 0.940 \\
200 & 1980 & 0.9 & IPW & EPD & 0.012 & 0.035 & 0.187 & 0.950 \\
200 & 1980 & 0.9 & IPW & GPD & 0.020 & 0.073 & 0.270 & 0.950 \\
500 & 4950 & 0.9 & IPW & EPD & -0.009 & 0.020 & 0.141 & 0.970 \\
500 & 4950 & 0.9 & IPW & GPD & -0.010 & 0.042 & 0.205 & 0.980 \\
1000 & 9990 & 0.9 & IPW & EPD & -0.002 & 0.014 & 0.119 & 0.920 \\
1000 & 9990 & 0.9 & IPW & GPD & 0.000 & 0.032 & 0.179 & 0.920 \\
200 & 1980 & 0.9 & AIPW & EPD & -0.004 & 0.009 & 0.094 & 0.950 \\
200 & 1980 & 0.9 & AIPW & GPD & -0.008 & 0.017 & 0.130 & 0.950 \\
500 & 4950 & 0.9 & AIPW & EPD & -0.003 & 0.005 & 0.066 & 0.960 \\
500 & 4950 & 0.9 & AIPW & GPD & -0.006 & 0.011 & 0.103 & 0.970 \\
1000 & 9990 & 0.9 & AIPW & EPD & -0.002 & 0.003 & 0.049 & 0.940 \\
1000 & 9990 & 0.9 & AIPW & GPD & -0.005 & 0.006 & 0.074 & 0.940 \\
\hline
\end{tabular}
\begin{tablenotes}
\footnotesize
\item \textbf{Abbreviations:} IPW, Inverse Probability Weighting; AIPW, Augmented IPW; EPD, Extrapolation Proportional Difference; GPD, Group Proportional Difference.
\end{tablenotes}
\end{threeparttable}
\end{table}

\section{Proof}
\subsection{Identification Proof}
The g-formula for identification under the non-nested design satisfying the positivity assumption is given by:

\begin{align*}
\mathbb{E}[Y(a) \mid S = 0] &= \mathbb{E}\left\{\mathbb{E}[Y(a) \mid \bX, S = 0] \right\} \\
                 &= \mathbb{E}\left\{\mathbb{E}[Y(a) \mid \bX, S = 0, D = 1] \mid S = 0, D = 1\right\}  \\
                 &= \mathbb{E}\left\{\mathbb{E}[Y(a) \mid \bX, S = 1, D = 1] \mid S = 0, D = 1\right\} \\
                 &= \mathbb{E}\left\{\mathbb{E}[Y(a) \mid A = a, \bX, S = 1, D = 1] \mid S = 0, D = 1 \right\} \\
                 &= \mathbb{E}\left\{\mathbb{E}[Y \mid A = a, \bX, S = 1, D = 1] \mid S = 0, D = 1\right\} \\
                 &= \mathbb{E}\left[\mathbb{E}[Y \mid A = a, \bX, S = 1, D = 1] \mid S = 0, D = 1 \right].
\end{align*}

The reweighting formula for identification under the non-nested design satisfying the positivity assumption is given by:

\vspace{-10 pt}
\allowdisplaybreaks
{\small
\begin{align*}
\mathbb{E}[Y(a) \mid S=0] 
&= \mathbb{E}\left\{\mathbb{E}[Y(a) \mid \bX, S=0]\right\} \\
&= \frac{1}{\Pr(S=0)} \mathbb{E}\left\{(1-S)\mathbb{E}[Y(a) \mid \bX, S=0]\right\} \\
&= \frac{1}{\Pr(S=0)} \mathbb{E}\left\{\Pr(S=0 \mid \bX)\mathbb{E}[Y(a) \mid \bX]\right\}  \\
&= \frac{1}{\Pr(S=0)} \mathbb{E}\left\{\Pr(S=0 \mid \bX) 
    \frac{\Pr(A=a \mid S=1, \bX)\Pr(S=1 \mid \bX)\mathbb{E}[Y(a) \mid \bX]}{\Pr(A=a \mid S=1, \bX)\Pr(S=1 \mid \bX)}\right\} \\
&= \frac{1}{\Pr(S=0)} \mathbb{E}\left\{\Pr(S=0 \mid \bX) 
    \frac{\mathbb{E}[I(A=a) \mid S=1, \bX]\mathbb{E}[I(S=1) \mid \bX]\mathbb{E}[Y(a) \mid \bX]}{\Pr(A=a \mid S=1, \bX)\Pr(S=1 \mid \bX)}\right\} \\
&= \frac{1}{\Pr(S=0)} \mathbb{E}\left\{\Pr(S=0 \mid \bX) 
    \frac{\mathbb{E}[I(A=a, S=1)Y(a) \mid \bX]}{\Pr(A=a \mid S=1, \bX)\Pr(S=1 \mid \bX)}\right\} \\
&= \frac{1}{\Pr(S=0)} \mathbb{E}\left\{\Pr(S=0 \mid \bX) 
    \frac{\mathbb{E}[I(A=a)I(S=1)Y \mid \bX]}{\Pr(A=a \mid S=1, \bX)\Pr(S=1 \mid \bX)}\right\}  \\
&= \frac{1}{\Pr(S=0 \mid D=1)} \mathbb{E}\left\{
    \frac{I(S=1)I(A=a)Y\Pr(S=0 \mid \bX, D=1)}{\Pr(S=1 \mid \bX, D=1)\Pr(A=a \mid \bX, S=1, D=1)}\right\}  \\
&= \frac{1}{\Pr(S=0 \mid D=1)} \mathbb{E}\left\{
    \frac{(1-h_s(\bX))I(S=1)I(A=a)Y}{h_s(\bX)e_a(\bX)}\right\},
\end{align*}
}

The identification proof under the positivity violation is given by:
{
\small
\begin{align*}
E(Y(a)|S=0) &= Pr(X \in \mathcal{R}_1|S=0)E(Y(a) \mid X \in \mathcal{R}_1, S=0) +  \\&Pr(X \in \mathcal{R}_2|S=0)E(Y(a) \mid X \in \mathcal{R}_2, S=0) + \\
&Pr(X \in \mathcal{R}_3|S=0)E(Y(a) \mid X \in \mathcal{R}_3, S=0)
\end{align*}
}

For \( E(Y(a) \mid X \in \mathcal{R}_3, S=0) \), where the positivity assumption holds, identification follows from the previously established weighting formula and g-formula. For \( E(Y(a) \mid X \in \mathcal{R}_1, S=0) \) when the positivity assumption is violated and \( E(Y(a) \mid X \in \mathcal{R}_2, S=0) \) when the positivity assumption is nearly violated, identification relies on the additional sensitivity assumption.

\subsection{Semiparametric Efficiency Bound}
To derive the semiparametric efficiency bound of $\tau$, we use the same approach as \citet{hahn1998role}, with theory developed by \citet{robins1995semiparametric} and \citet{robins1995analysis}.

The probability density of the observed data, denoted by $\bm o$, under the assumptions and biased sampling framework, is expressed as:
\begin{align*}
q(\bm o) &= q(s)p(\bx|s)\left[p(a|\bx)p(y|a,\bx)\right]^s,
\end{align*}
where $q(s)$ represents the probability of trial sampling under biased sampling, $p(\bx|s)$ denotes the probability density of $\bx$ conditional on the group $s$, $p(a|\bx)$ is the probability of treatment $a$ conditional on $\bx$, and $p(y|a, \bx)$ is the probability density of $y$ given treatment $a$ and $\bx$.

Let $q(\bm o;\bm{\gamma})$ be a parametric submodel with parameter $\bm{\gamma}$ and $q(\bm o;\bm{\gamma}) = q(\bm{o};\bm{\gamma_0})$.

Based on the g-formula identification result, we write
$$\tau(\bm\gamma) =  \iint y[p(y|\bx, a = 1;\bm\gamma)-p(y|\bx, a = 0;\bm\gamma)]p(\bx|s = 0;\bm \gamma)dyd\bx.$$

By definition, the efficient influence function under biased sampling distribution $Q$ is the unique function $\varphi(\bm O)$ that satisfies $\partial\tau(\bm\gamma)/\partial\bm\gamma \mid_{\bm \gamma = \bm \gamma_0} =E_Q[\varphi(\bm O)G_{\bm \gamma}(\bm{O})]$, where $G_{\bm \gamma}(\bm{O})$ is defined as $\partial \log q(\bm o; \bm \gamma)/\partial\bm\gamma \mid_{\bm \gamma = \bm \gamma_0}$ with
$$
\frac{\partial \log q(\bm o; \bm \gamma)}{\partial\bm \gamma}\bigg|_{\bm \gamma = \bm{\gamma_0}} = s\cdot G_y(y, a, \bx) + s\cdot G_a(a,\bx) + G_x(\bx, s) + G_s(s),
$$
where $G_y(y, a, \bx) = \frac{\partial}{\partial \bm \gamma} p(y \mid \bx, a; \bm \gamma) \big|_{\bm \gamma = \bm{\gamma_0}}$, $G_a(a, \bx) = \frac{\partial}{\partial \bm \gamma} \log p(a \mid \bx; \bm \gamma) \big|_{\bm \gamma = \bm{\gamma_0}}$, $G_x(\bx, s) = \frac{\partial}{\partial \bm \gamma} \log p(\bx \mid s; \bm \gamma) \big|_{\bm \gamma = \bm{\gamma_0}}$, and $G_s(s) = \frac{\partial}{\partial \bm \gamma} \log q(s; \bm \gamma) \big|_{\bm \gamma = \bm{\gamma_0}}$.

Then, it is straightforward to show that $\partial\tau(\bm\gamma)/\partial\bm\gamma \mid_{\bm \gamma = \bm \gamma_0}$ equals
$$
E\left\{ E\bigg[Y\big(G_{y}(Y,1,\bX) + G_{x}(\bX,0)\big) \mid A = 1, \bX\bigg] 
- E\bigg[Y\big(G_{y}(Y,0,\bX) + G_{x}(\bX,0)\big) \mid A = 0, \bX\bigg] \Big| S=0 \right\}.
$$

Denote the function as
\begin{align*}
\varphi (\bm O) &= \frac{1}{q(s=1)} \left\{ 
\frac{I(S=1)p(\bx \mid s = 0)}{p(\bx \mid s=1)} \left[ 
\frac{I(A=1)}{p(a=1 \mid \bX)} (Y - \mu_1(\bX)) 
- \frac{I(A=0)}{p(a=0 \mid \bX)} (Y - \mu_0(\bX)) 
\right]  \right\} \\
&\quad + \frac{I(S=0)}{q(s=0)} \tau(\bX) \\
&= \frac{1}{q(s=0)} \left\{ 
\frac{I(S=1)(1-h_s(\bX))}{h_s(\bX)} \left[ 
\frac{I(A=1)}{e_1(\bX)} (Y - \mu_1(\bX)) 
- \frac{I(A=0)}{e_0(\bX)} (Y - \mu_0(\bX)) 
\right] + I(S=0) \tau(\bX)  \right\}.
\end{align*}

It is easy to show this choice of $\varphi(\bm{O})$ satisfies $\partial\tau(\bm\gamma)/\partial\bm\gamma \mid_{\bm \gamma = \bm \gamma_0} =E_Q[\varphi(\bm O)G_{\bm \gamma}(\bm{O})]$. Thus, $\varphi(\bm O)$ is the efficient influence function for $\tau$. The semiparametric efficiency bound is then the expected square of the efficient influence function, given by 
{\small
\[
V(\tau) = \frac{1}{(q(s=0))^2}\mathbb{E}\left[\left\{\frac{(1-h_s(\bX))^2}{h_s(\bX)}\right\}\left(\frac{\sigma_1^2(\bX)}{e_1(\bX)} + \frac{\sigma_0^2(\bX)}{e_0(\bX)}\right) + (1-h_s(\bX))(\tau(\bX) - \tau)^2\right].
\]
}

\subsection{Proof of Asymptotic Properties}
The asymptotic normality of the weighting estimator and the augmented weighting estimator follows from M-estimation theory \cite{van2000asymptotic}.

First, note that under correct parametric model specification:
$$
E[\bm\Psi_w(S, \bm{X}, A, Y; \bm{\Xi}{w,0})] = \bm{0}, \quad E[\bm\Psi{aw}(S, \bm{X}, A, Y; \bm{\Xi}_{aw,0})] = \bm{0}.
$$

These equations hold due to the standard properties of score functions, the definition of the threshold satisfying the proportion equation, the reweighting formula, and the g-formula. Thus, $\hat{\bm{\Xi}}_w$ and $\hat{\bm{\Xi}}_{aw}$ are the corresponding M-estimators. Under regularity conditions:

$$
\sqrt{n}(\hat{\bm{\Xi}}_w - \bm{\Xi}_w) \xrightarrow{d} N\left(0, \bm{A}_w^{-1} \bm{B}_w \bm{A}_w^{-\top}\right),
$$
$$
\sqrt{n}(\hat{\bm{\Xi}}_{aw} - \bm{\Xi}_{aw}) \xrightarrow{d} N\left(0, \bm{A}_{aw}^{-1} \bm{B}_{aw} \bm{A}_{aw}^{-\top}\right).
$$

The asymptotic normality of $\hat{\tau}_{3w}$ and $\hat{\tau}_{3aw}$ is then derived using the delta method.

\bibliographystyle{abbrvnat} 
\bibliography{sample}